\documentclass{article}
\usepackage[pdftex]{graphicx}
\usepackage{float}

\begin{document}

{\Large Branching stochastic processes as models}

{\Large of Covid-19 epidemic development\bigskip }

\textbf{Nikolay M. Yanev}\bigskip $^{1}$, \textbf{Vessela K. Stoimenova}$%
^{2} $\textbf{, Dimitar V. Atanasov}$^{3}$\bigskip

\textbf{Abstract.}

The aim of the paper is to describe two models of Covid-19 infection
dynamics. For this purpose a special class of branching processes with two
types of individuals is considered. These models are intended to use only
the observed daily statistics to estimate the main parameter of the
infection and to give a prediction of the mean value of the non-observed
population of the infected individuals. Similar problems are considered also
in the case when the processes admit an immigration component. This is a
serious advantage in comparison with other more complicated models where the
officially reported data are not sufficient for estimation of the model
parameters. In this way the specific development of the Covid-19 epidemics
is considered also for all countries as it is given in the specially created
site http://ir-statistics.net/covid-19 where the obtained results are
updated daily.\bigskip \bigskip

MSC-2020: Primary 92D30

Secondary 60J80; 60J85; 62P10

\bigskip , branching processes,
immigration, modeling, estimation. 

\textbf{1. Introduction.}

The theory of branching processes is a powerful tool for investigation the
population dynamics where the members can reproduce new members following
some stochastic laws. The objects may be of different types and nature.
Branching processes have serious applications in physics, chemistry, biology
and medicine, demography, epidemiology, economics, computer science an so
on. Basic models and analytical results are presented in some books and a
lot of papers. We would like to point out the monographs $[1-8]$ among the
others. Some applications of branching processes in biology and medicine are
presented in $[5]$, $[9]$ and $[10]$. \ For statistical inference of
branching processes one can consider $[11]$ and $[12].$ Some specific
estimation problems are given in $[13]$, $[14]$ and $[15]$.

The aim of the present paper is to model and to estimate the development of
the Covid-19 infection in the population. For this purpose a special class
of branching processes with two types of infected individuals is constructed
and considered day by day. In fact they are "infected undiagnosed" vs "
infected diagnosed" following the terminology of P. Jagers (personal
communication). It is proposed also a generalization of this situation
assuming an immigration component. In this way we are able to use the
observed data for the Covid-19 daily registered infected individuals and to
estimate the main parameter of infection. In fact this parameter $m$\
represents the mean value of the infected individuals by one individual per
day. Using the observed statistics some methods for estimation are proposed
and corresponding graphics are presented. Two models with and without
immigration are compared. In this way we are able to give a prediction of
the possible development of the mean value of the infected individuals.

Notice that both type processes with or without immigration have an
exponential growth in the supercritical case $m>1$ but in the critical case $%
m=1$ and in the subcritical case $m<1$ the asymptotic behaviour is
essentially different. In the critical case the mean value of the process
with immigration grows linearly while for the process without immigration
the mean value is constant. In the subcritical case the mean value of the
immigration process converges to a positive constant but for the process
without immigration the mean value goes to zero. The estimation of the
immigration mean $a$ is a serious problem because this parameter cannot be
estimated by the observed data and one needs additional information. Hence
the processes with immigration need more careful investigation.

As it is given in the paper the proposed estimators can be applied also in
the case when the processes are inhomogeneous in time. The behaviour of the
estimators shows that the observed processes are able to change the
criticality during the development of the epidemics. The estimated values
for the main epidemical parameter $m$ vary greater than $1$, equal (or very
closed) to $1$ and even less than $1.$ Moreover it seems that the real
epidemic process develops like a mixture of both type of the models with and
without immigration. In this case four stages of epidemic development are
available: exponential growth ($m>1$), linear growth ($m=1$ and an
immigration component), non-increasing and almost stable population (due to $%
m=1$ or $m<1$ with an immigration component), convergence to zero
(extinction of the epidemics due to $m<1$ without immigration). It is
obvious that the restriction of the immigration component is very important
to the limitation of the epidemic process. It seems that in some countries
exist some regions which can be considered as immigration sources for the
other regions and in general for the whole country. The abroad immigration
plays also an important role.

The paper continues the investigations started in $[16]$ where some results
for the model without immigration were presented only for Bulgaria, Italy
and globally.

In the present paper as an illustration of the models with and without
immigration the obtained results are presented for several countries: USA,
Italy, France, Germany, Spain and Bulgaria. Additional information, reports
and plots, related to this research for all countries all over the world can
be found on the site {http://ir-statistics.net/covid-19. The data used for
the estimation of the parameters of the model are taken from European Centre
for Disease Prevention and Control $[21]$, similar to the data provided by
World Health Organization $[20]$. Since these databases are updated daily,
the proposed here model is applied regularly on each new data set. Using
these results one can compare the infection rate on different countries and
regions on the basis of estimated growth rate. For example, on Table \ref%
{tab:estComp}, the 10 countries with lowest and highest growth rate are
shown. Even in the cases where the infection growth is less than 1, the
confidence interval goes above 1, which states that there is a possibility
for increasing of the infection growth in the future. }

The theoretical model based on two type branching process is described in
detail in Section 2. The p.g.f.'s and the mathematical expectations are
obtained. Regardless of its simplicity the model has a great advantage using
only the observed official data for the lab-confirmed cases. The two-type
branching process assuming an additional immigration component is considered
in Section 3. The estimation problems are presented in Section 4. Some
conclusive remarks are given in Section 5.

Finally the estimation of the mean parameter of infection can be considered
as a fast test to estimate the rate of Covid-19 epidemic in a country or a
region. It can be used also as a first stage of a construction to some more
complicated epidemiological models where it is not possible to estimate
directly this parameter. Obviously the solution of this problem requires the
collaboration of specialists in various fields as epidemiology, mathematics,
medicine, microbiology, molecular biology and informatics among others.

\textbf{2 Two-type branching process as model of Covid-19 population
dynamics.}

Assume that the epidemic process of infection begins with some finite number
of immigrants and then the process of immigration is isolated under the
quarantine.

To describe this situation we can consider a two type branching process $%
\{Z_{1}(n),Z_{2}(n)\}$ where type $T_{1}$ are infected (but still healthy)
individuals who don't know that they are Covid-19 infected and type $T_{2}$
of discovered with Covid-19 virus individuals (and this is the data we use).
Every individual of type $T_{1}$ (infected) produces per day a random number
of new individuals of type $T_{1}$ (infected) or only one individual of type 
$T_{2}$ (more precisely, in this case the individual type $T_{1}$ is
transformed into an individual type $T_{2}).$ Note that $T_{2}$ is a final
type, i.e. the individuals of this type don't take part in the further
evolution of the process because they are isolated under the quarantine.

Let $\mathbf{\xi }_{1}=(\xi _{1}^{(1)},\xi _{2}^{(1)})$ be the offspring
vector of type $T_{1}.$ Then the offspring joint probability generating
function (p.g.f.) of type $T_{1}$ can be defined as follows:\bigskip \newline
$(1)$ $\ \ h_{1}(s_{1},s_{2})=\mathbf{E(}s_{1}^{\xi _{1}^{(1)}}s_{2}^{\xi
_{2}^{(1)}})=p_{0}+\sum_{j=1}^{k}p_{j}s_{1}^{j}+qs_{2},|$ $s_{1}|\leq
1,|s_{2}|\leq 1,$\bigskip \newline
where $q=1-\sum_{j=0}^{k}p_{j},$ $h_{1}(1,1)=1.$

Obviously $h_{2}(s_{1},s_{2})\equiv 1$ because the type $T_{2}$ has $(0,0)$
offspring.

Note that $p_{0}$ is the probability that type $T_{1}$ goes out of the
reproduction process (the individual becomes healthy or goes out of the
country, i.e. emigrates), $p_{j}$ is the probability to produce new $j$
infected individuals of type $T_{1}$ and $q$ is the probability that the
individual type $T_{1}$ is confirmed ill (or dead). In other words, $q=%
\mathbf{P}\{T_{1}\rightarrow T_{2}\},$ i.e. with probability $q$ an
individual of type $T_{1}$ is transformed into an individual of type $T_{2}.$
Then from $(1)$ we can obtain also that the marginal p.g.f. are\newline
$(2)$ $\ \ \ \mathbf{E(}s_{1}^{\xi
_{1}^{(1)}})=h_{1}(s_{1},1)=p_{0}+\sum_{j=1}^{k}p_{j}s_{1}^{j}+q=1-%
\sum_{j=1}^{k}p_{j}(1-s_{1}^{j}),$\newline
$(3)$ $\ \ \ \mathbf{E(}s_{2}^{\xi _{2}^{(1)}})=h_{1}(1,s_{2})=1-q+qs_{2}.$

If we assume that $Z_{1}(0)>0$ and $Z_{2}(0)=0$ then for $n=1,2,...$ \newline
$(4)$ \ \ \ $Z_{1}(n)=\sum_{j=1}^{Z_{1}(n-1)}\xi _{1}^{(1)}(n;j),$ $%
Z_{2}(n)=\sum_{j=1}^{Z_{1}(n-1)}\xi _{2}^{(1)}(n;j),$ \newline
where \ the vectors $\{(\xi _{1}^{(1)}(n;j),\xi _{2}^{(1)}(n;j)\}$ are
independent and identically distributed (iid) as $(\xi _{1}^{(1)},\xi
_{2}^{(1)}).$

The recurrent formula $(4)$ defines two-type branching process $%
\{(Z_{1}(n),Z_{2}(n)),n=0,1,2,...\}.$ Notice that\textit{\ }$Z_{1}(n)$ is
the total number of individuals (type $T_{1}$) in the $n$-th day infected by
the individuals of the $(n-1)$-th day; $Z_{2}(n)$ is the total number of the
registered Covid-19 individuals (type $T_{2}$) in the $n$-th day. The
process starts with $Z_{1}(0)$ infected individuals, where $Z_{1}(0)$ can be
an integer-valued random variable with a p.g.f. $h_{0}(s)=Es^{Z_{0}}=%
\sum_{k=1}^{K}p_{0k}s^{k},$ $|s|\leq 1$, or $Z_{0}=N$ for some integer
value, $N=1,2,...$ . The random variable $\xi _{1}^{(1)}(n;j)$ is the number
of individuals of type $T_{1}$ in the $n$-th day infected by the $j$-th
individual of type $T_{1}$ from the $(n-1)$-th day, $j=1,2,...,Z_{1}(n-1)$.
Similarly the random variable $\xi _{2}^{(1)}(n;j)$ is the number of the
confirmed infected individuals (type $T_{2})$ in the $n$-th day transformed
by the $j$-th infected individual type $T_{1}$ from the $(n-1)$-th day, $%
j=1,2,...,Z_{1}(n-1)$.

Note that $\mathbf{P}\{\xi _{2}^{(1)}(n;j)=0\}=1-q$ and $\mathbf{P}\{\xi
_{2}^{(1)}(n;j)=1\}=q.$ Hence $Z_{2}(n)\in Bi(Z_{1}(n-1,q),$ i.e. \newline
$\mathbf{P}\{Z_{2}(n)=i|Z_{1}(n-1)=l%
\}=(_{i}^{l})q^{i}(1-q)^{l-i},i=0,1,...,l;l=0,1,2,...$

In other words the probability $q$ can be interpreted as a proportion of the
confirmed individuals in the day $n$ among all infected individuals in the
day $n-1$.

Let $h_{0}(s)=\mathbf{E}s^{Z_{1}(0)},$ $F_{1}(n;s)=\mathbf{E(}s^{Z_{1}(n)}),$
$F_{2}(n;s)=\mathbf{E(}s^{Z_{2}(n)}).$\ Introduce the following p.g.f.%
\newline
$(5)$ $\ \ \ h^{\ast }(s)=h_{1}(s,1)=q+p_{0}+\sum_{j=1}^{k}p_{j}s^{j},$ $%
\widetilde{h}(s)=h_{1}(1,s)=1-q+qs.$

Then it is not difficult to check that for $n=0,1,2,...,$ we are able to
obtain the p.g.f. of the process:\newline
$(6)$ $\ \ \ F_{1}(n;s)=\mathbf{E(}s^{Z_{1}(n)})=F_{1}(n-1;h^{\ast
}(s))=F_{1}(0;h_{n}^{\ast }(s))$

$\ \ \ \ \ \ \ \ \ \ \ \ \ \ =h_{0}(h^{\ast }(h^{\ast }(...(h^{\ast
}(s))...))),$\newline
$(7)$ $\ \ \ F_{2}(n;s)=\mathbf{E(}s^{Z_{2}(n)})=F_{1}(n-1;\widetilde{h}%
(s))=F_{1}(0;\widetilde{h}_{n}(s))$

$\ \ \ \ \ \ \ \ \ \ \ \ \ \ =h_{0}(\widetilde{h}(\widetilde{h}(...(%
\widetilde{h}(s))...))),\newline
$where the p.g.f. $h_{n}^{\ast }(s)$ and\ $\widetilde{h}_{n}(s)$ are
obtained after $n$ compositions of the p.g.f. $h^{\ast }(s)$ and $\widetilde{%
h}(s)$ \newline
$(8)$ $\ \ h_{n}^{\ast }(s)=h^{\ast }(h^{\ast }(...(h^{\ast
}(s))...)),h_{0}^{\ast }(s)=s;\widetilde{h}_{n}(s)=\widetilde{h}(\widetilde{h%
}(...(\widetilde{h}(s))...)),\widetilde{h}_{0}(s)=s.$

Let $m=\frac{d}{ds}h^{\ast }(s)|_{s=1}=\mathbf{E}\xi
_{1}^{(1)}=\sum_{j=1}^{k}jp_{j}$ be the mean value of the new infected
individuals by one infected individual. Note that $\frac{d}{ds}\widetilde{h}%
(s)|_{s=1}=\mathbf{E}\xi _{2}^{(1)}=q$ is the mean value of the registered
infected individuals by one infected individual. Introduce also $m_{0}=%
\mathbf{E}Z_{1}(0)=\frac{d}{ds}h_{0}(s)|_{s=1}.$ Therefore \newline
$(9)$ \ \ \ $\ \ M_{1}(n)=\mathbf{E}Z_{1}(n)=m_{0}m^{n},n=0,1,2,...,$ 
\newline
$(10)$ \ $\ \ M_{2}(n)=\mathbf{E}Z_{2}(n)=q\mathbf{E}%
Z_{1}(n-1)=qm_{0}m^{n-1},n=1,2,...;\mathbf{E}Z_{2}(0)=0.$

Notice that the asymptotic behaviour of the process depends essentially of
parameter $m.$ Especially, if $m>1$ (supercritical case) then the mean value
of the infected individuals $M_{1}(n)$ grows exponentially, in the critical
case $m=1$ it is a constant and for $m<1$ (subcritical case) $%
M_{1}(n)\rightarrow 0$ as $n\rightarrow \infty .$

We will use these results to present in the next section a more complicated
model with immigration.

Note that we can observe only $Z_{2}(1),Z_{2}(2),...,Z_{2}(n).$ What can be
estimated with these observations?

Note first that $\frac{\mathbf{E}Z_{2}(n+1)}{\mathbf{E}Z_{2}(n)}=m.$ Hence
we can consider \newline
$(11)$ $\ \ \ \ \ \widehat{m}_{n}=\frac{Z_{2}(n+1)}{Z_{2}(n)},n=1,2,....,$ 
\newline
as an estimator of the parameter $m$ (similar to Lotka-Nagaev estimator for
the classical BGW branching process). It is possible to use also the
following Harris type estimator \newline
$(12)$ \ \ \ \ $\widetilde{m}_{n}=\sum_{i=2}^{n+1}Z_{2}(i)/%
\sum_{j=1}^{n}Z_{2}(j),n=1,2,....,$ \newline
or Crump and Hove type estimators \newline
$(13)$ \ \ \ \ $\overline{m}_{n,N}=\sum_{i=n+1}^{n+N}Z_{2}(i)/%
\sum_{j=n}^{n+N-1}Z_{2}(j),n=1,2,...;N=1,2,....$

See $[12]$ for more details.

Estimating $m$ we are able to predict the mean value of the infected (non
observed) individuals in the population. In the case when we assume that $%
Z_{1}(0)=1$ then $M_{1}(n)=\mathbf{E}Z_{1}(n)$ can be approximated
respectively by\ $\widehat{m}_{n}^{n}$, or $\widetilde{m}_{n}^{n}$, or $%
\overline{m}_{n,N}^{n}.$ In fact it means that we can obtain three types of
estimators \newline
$(14)$ \ \ \ $\widehat{M}_{1}(n)=$\ $\widehat{m}_{n}^{n},$ $\widetilde{M}%
_{1}(n)=$\ $\widetilde{m}_{n}^{n}$ and $\overline{M}_{1}(n)=\overline{m}%
_{n,N}^{n}.$

In other words we could say that we have at least three prognostic lines.
Therefore if we have the observations $(Z_{2}(1),Z_{2}(2),...,Z_{2}(n))$
over the first $n$ days, we are able to predict the mean value of the
infected individuals for the next $k$ days by the relations: \newline
$(15)$ \ \ \ $\widehat{M}_{1}(n+k)=$\ $\widehat{m}_{n}^{n+k},$ $\widetilde{M}%
_{1}(n+k)=$\ $\widetilde{m}_{n}^{n+k}$ and $\overline{M}_{1}(n+k)=\overline{m%
}_{n,N}^{n+k},$ $k=1,2,...$

We are able to estimate also the proportion $\alpha (n)$ of the registered
infected individuals among the population in the $n$-th day. Then we can
obtain the following three types of estimators: $\widehat{\alpha }(n)=$ $%
Z_{2}(n)/\{Z_{2}(n)+\widehat{M}_{1}(n)\},$ $\widetilde{\alpha }%
(n)=Z_{2}(n)/\{Z_{2}(n)+\widetilde{M}_{1}(n)\},$ $\overline{\alpha }%
(n)=Z_{2}(n)/\{Z_{2}(n)+\overline{M}_{1}(n)\}.$

All obtained estimators will be presented by the observed registered
lab-confirmed cases. The quality of the estimation, however, depends on the
representativeness of the sample due to the specifics of the data collection
in each country.

\textbf{Remark 1. }In fact our model $(4)$ can be generalized as
non-homogeneous in time. In this case $m(l)=\frac{d}{ds}h^{\ast
}(l;s)|_{s=1}=\mathbf{E}\xi _{1}^{(1)}(l)=\sum_{j=1}^{k}jp_{j}(l)$ will be
the mean value of the new infected individuals by one infected individual in
the day $l=1,2,...$. Therefore instead $(9)$ we obtain \newline
$(16)$ \ \ \ $\ \ M_{1}(n)=\mathbf{E}Z_{1}(n)=m_{0}\Pi
_{l=1}^{n}m(l),n=0,1,2,....$ \newline

Notice that in this case $\frac{\mathbf{E}Z_{2}(n+1)}{\mathbf{E}Z_{2}(n)}%
=m(n).$ Hence we can use $(11)-(13)$ to estimate $M_{1}(n)$ from $(16).$
Therefore \newline
$(17)$ \ \ \ \ $\widehat{M}_{1}(n)=\Pi _{l=1}^{n}$\ $\widehat{m}_{l},$ $%
\widetilde{M}_{1}(n)=$\ $\Pi _{l=1}^{n}\widetilde{m}_{l},n=1,2,..$

\textbf{3. The two-type branching process with immigration as model of
Covid-19 epidemic development.}

The model considered in Section 2 assumes that the process of infection
begins with some random number of infected immigrants and then the
immigration process is bounded and it is not essential for the Covid-19
population dynamics. But in some cases the role of the immigration process
cannot be negligible. That is why we will introduce random variables $%
\{I_{n}\},$ where $I_{n}$ gives the number of infected immigrants in the $%
n-th$ day which take part in the process of infection. We will assume first
that $\{I_{n}\}$ are iid r.v. with a p.g.f. \newline
$(18)\ \ \ $ \ \ $g(s)=\mathbf{E}s^{I_{n}}=\sum_{k=0}^{l}g_{k}s^{k},|s|\leq
1.$

Then instead of $(4)$ we will consider the following branching process with
immigration \newline
$(19)$ \ \ \ $Y_{1}(n)=\sum_{j=1}^{Y_{1}(n-1)}\xi _{1}^{(1)}(n;j)+I_{n},$ $%
Y_{2}(n)=\sum_{j=1}^{Y_{1}(n-1)}\xi _{2}^{(1)}(n;j),n=1,2,...,$ \newline
where \ the vectors $\{(\xi _{1}^{(1)}(n;j),\xi _{2}^{(1)}(n;j)\}$ are
independent and identically distributed as $(\xi _{1}^{(1)},\xi _{2}^{(1)})$
with p.g.f. $(1)-(3)$ and they are also independent of $\{I_{n}\}.$ We can
assume that $Y_{1}(0)>0$ is some random variable independent of $\{(\xi
_{1}^{(1)}(n;j),\xi _{2}^{(1)}(n;j)\}$ and $\{I_{n}\},$ and also the case$\
Y_{2}(0)=0.$ Another possible assumption is $Y_{1}(0)=Y_{2}(0)=0,$ which
means that in fact the process starts with the first real immigrants.

\textit{Interpretation: }$Y_{1}(n)$ is the total number of individuals (type 
$T_{1}$) in the $n$-th day infected by the individuals of the $(n-1)$-th day
plus the new infected immigrants $I_{n}$; $Y_{2}(n)$ is the total number of
the officially registered infected individuals (type $T_{2}$) in the $n$-th
day.

Then it is not difficult to check that for $n=0,1,2,...,$ we are able to
obtain from $(18)$ and $(19)$ the p.g.f.'s of the process:\newline
$\ \ \ \ \ \ G_{1}(n;s)=\mathbf{E(}s^{Y_{1}(n)})=g(s)G_{1}(n-1;h^{\ast
}(s))=f(h_{n}^{\ast }(s))\Pi _{k=0}^{n-1}g(h_{k}^{\ast }(s)),$\newline
$\ \ \ \ \ \ G_{2}(n;s)=\mathbf{E(}s^{Y_{2}(n)})=G_{1}(n-1;\widetilde{h}(s))$

$\ \ \ \ \ \ \ \ \ \ \ \ \ =g(\widetilde{h}(s))G_{1}(n-2;h^{\ast }(%
\widetilde{h}(s)))=f(h_{n-2}^{\ast }(\widetilde{h}(s)))\Pi
_{k=0}^{n-1}g(h_{k}^{\ast }(\widetilde{h}(s))),\newline
$where the p.g.f. $h_{k}^{\ast }(s)$ and\ $\widetilde{h}_{k}(s)$ are
obtained after $k$ iterations of the p.g.f. $h^{\ast }(s)$ and $\widetilde{h}%
(s)$ as it is given in $(8)$ and \newline
$\ $ \ \ \ \ \ $f(s)=\mathbf{E}s^{Y_{1}(0)}=\sum_{k=0}^{N}f_{k}s^{k},|s|\leq
1.$

Notice that if we assume that $Y_{1}(0)=0$ then $f(s)\equiv 1$ and \newline
$(20)$ \ \ \ \ $G_{1}(n;s)=\Pi _{k=0}^{n-1}g(h_{k}^{\ast }(s)),$ $%
G_{2}(n;s)=\Pi _{k=0}^{n-2}g(h_{k}^{\ast }(\widetilde{h}(s))).$

From $(18)$ we can introduce the immigration mean $a=\mathbf{E}I_{n}=\frac{d%
}{ds}g(s)|_{s=1}=\sum_{k=0}^{l}kg_{k}$ . Then from \ $(18)-(20)$ it is not
difficult to obtain that\ for $n=1,2,...$\ \newline
$(21)$ \ \ \ \ $A_{1}(n)=\mathbf{E}Y_{1}(n)=mA_{1}(n-1)+a=a%
\sum_{k=0}^{n-1}m^{k},$\newline
$(22)$ \ \ \ $A_{2}(n)=\mathbf{E}Y_{2}(n)=qA_{1}(n-1)=qa%
\sum_{k=0}^{n-2}m^{k},$ \newline
where it is assumed that $A_{1}(0)=\mathbf{E}Y_{1}(0)=0$ and the parameters $%
m$ and $q$ are well defined in Section 2 by $(5).$

Hence from $(21)$ and $(22)$ one has \newline
$(23)$ \ \ \ $A_{1}(n)=a(m^{n}-1)/(m-1),m\neq 1,$ and $A_{1}(n)=an,m=1,$%
\newline
$(24)$ \ \ \ $A_{2}(n)=qa(m^{n-1}-1)/(m-1),m\neq 1,$ and $%
A_{2}(n)=qa(n-1),m=1$

Therefore by $(23)$ and $(24)$ one obtains as $n\rightarrow \infty $\newline
$A_{1}(n)\sim am^{n}/(m-1),m>1;A_{1}(n)=an,m=1;A_{1}(n)\rightarrow
a/(1-m),m<1,$\newline
$A_{2}(n)\sim qam^{n-1}/(m-1),m>1;A_{2}(n)\sim qan,m=1;A_{1}(n)\rightarrow 
\frac{qa}{1-m},m<1.$

In the general case $A_{1}(0)=\mathbf{E}Y_{1}(0)=M_{0}>0$ and instead of $%
(19)$ and $(20)$ one has \newline
$(25)$ \ \ \ $A_{1}(n)=M_{0}m^{n}+a\sum_{k=0}^{n-1}m^{k},$\newline
$(26)$ \ \ \ $A_{2}(n)=q(M_{0}m^{n-1}+a\sum_{k=0}^{n-2}m^{k}).$

We would like to point out once again that we can observe only the
statistics $Y_{2}(1),Y_{2}(2),...,Y_{2}(n)$ and we have to use for
estimation only these observations.

Notice first that for $m\geq 1$ we obtain $\lim_{n\rightarrow \infty }\frac{%
\mathbf{E}Y_{2}(n+1)}{\mathbf{E}Y_{2}(n)}=m.$ Hence for large enough $n$ we
can consider \newline
$(27)$ $\ \ \ \ \ \widehat{m}_{n}=Y_{2}(n+1)/Y_{2}(n)$\newline
as an estimator of the parameter $m$ (similar to Lotka-Nagaev estimator for
the classical BGW branching process). It is possible to use also for $m>1$
and large enough $n$ the following Harris type estimator \newline
$(28)$ \ \ \ \ $\widetilde{m}_{n}=\sum_{i=2}^{n+1}Y_{2}(i)/%
\sum_{j=1}^{n}Y_{2}(j)$\newline
or Crump and Hove type estimators \newline
$(29)$ \ \ \ \ $\overline{m}_{n,t}=\sum_{i=n+1}^{n+t}Y_{2}(i)/%
\sum_{j=n}^{n+t-1}Y_{2}(j).$

See $[12]$ for more details.

Estimating $m>1$ we are able to predict the mean value of the infected (non
observed) individuals in the population. In the case when we assume that the
process begins with the first immigrants then $A_{1}(n)=\mathbf{E}Y_{1}(n)$
can be approximated using the estimators\ $(27)-(29).$

The problem is how to estimate the immigration mean $a$. First of all there
is an special case when $a=m$. Then using $(23)$ and $(24)$ with the Harris
estimator we have\newline
$(30)$ \ \ $\widetilde{A}_{1}(n)=\widetilde{m}_{n}(\widetilde{m}_{n}^{n}-1)/(%
\widetilde{m}_{n}-1),m>1;$ $\widetilde{A}_{1}(n)=\widetilde{m}_{n}n,m=1.$

In general we have to use some additional information. For example, if we
can observe $\{I_{k}\}$ then we can apply the estimator $a_{n}^{\ast
}=n^{-1}\sum_{k=1}^{n}I_{k}.$ Hence \newline
$(31)$ \ \ \ \ \ \ \ \ \ \ $\widetilde{A}_{1}(n)=a_{n}^{\ast }(\widetilde{m}%
_{n}^{n}-1)/(\widetilde{m}_{n}-1),m>1;$ $\widetilde{A}_{1}(n)=a_{n}^{\ast
}n,m=1.$

One can proceed similarly for the other \ estimators $\widehat{m}_{n}^{n}$
and $\overline{m}_{n,N}^{n}.$

\textbf{Remark 2. }Similarly as it is shown in\textbf{\ Remark 1 }from
Section 2 the model $(19)$ can be generalized in the case with
non-homogeneous in time offspring distributions. In this case $m(l)=\frac{d}{%
ds}h^{\ast }(l;s)|_{s=1}=\mathbf{E}\xi _{1}^{(1)}(l)=\sum_{j=1}^{k}jp_{j}(l)$
will be the mean value of the new infected individuals by one infected
individual in the day $l=1,2,...$. Therefore instead $(25)$ we obtain 
\newline
$(32)$ \ \ \ $\ A_{1}(n)=\mathbf{E}Y_{1}(n)=M_{0}\Pi
_{l=1}^{n}m(l)+a\sum_{k=0}^{n-1}\Pi _{l=1}^{k}m(l).$

\textbf{3. Estimating of the main parameter and some predictions.}

Recall that both type processes with or without immigration have exponential
growth in the supercritical case $m>1$. In the critical case $m=1$ and in
the subcritical case $m<1$ the asymptotic behaviour is essentially
different. In the critical case the mean value of the process with
immigration grows linearly while for the process without immigration the
mean value is constant. In the subcritical case the mean value of the
immigration process converges to a positive constant but for the process
without immigration the mean value limit is equal to zero. The estimation of
the immigration mean $a$ cannot be estimated by the observed statistics and
we need some additional information.

We would like to point out once again that the considered in Section 2 model
is versatile but the application in each country is specific because it
depends essentially on the official data from the country. The plots and
tables below illustrate well some specific details for different countries
as well as the common trend.

The data used for the estimation of the parameters of the model come from
European Centre for Disease Prevention and Control $[21]$.

We will consider first the process without immigration. Note that the
observed data is the number of the newly (daily) registered individuals
denoted by $Z_{2}(n)$. The data about the new number of infected individuals
(denoted by $Z_{1}(n)$) is unobservable. The initial number $m_{0}=EZ_{1}(0)$
is also unknown. Here $n$ is the corresponding day from the beginning of the
infection.

The estimation of the parameters of the defined model can be summarized in
the following steps.

\begin{enumerate}
\item On the basis of each sample $Z_{2}(1),\dots ,Z_{2}(s)$, $s=1,\dots ,n$%
, the mean numbers of the new infected individuals by one infected
individual is estimated by the considered above estimators but we present
only the results for Harris type estimator.

\item The mean values of the expected number of nonregistered infected
individuals are calculated for the Harris estimator as $%
M_{1}(s+k)=m_{0}m^{s+k}=M_{1}(s)m^{k}$. Here, instead of $m_{0}$ the value
of $M_{1}(s)$ is estimated by the registered contaminated individuals in day 
$s$. For the purpose of the study, the value of $s$ is set to 20 days before
the end of observed data, i.e. $s=n-20$.

\item The proportion $\alpha (n)$ of the registered contaminated individuals
among the population of all infected in the $n$-th day is estimated by the
formula $\widetilde{\alpha }(n)=Z_{2}(n)/\{Z_{2}(n)+\widetilde{M}_{1}(n)\}$.

\item The expected number of individuals in the model with immigration $%
A_{1}(n)=\mathbf{E}Y_{1}(n)$ is calculated using equation $(25)$, based on
the Harris estimator, calculated above.

\item The obtained results are presented with 95 \% confident intervals.
\end{enumerate}

Firstly, we will demonstrate the approach described above by the data of the
reported laboratory-confirmed COVID-19 daily cases for USA provided by the
European Centre for Disease Prevention and Control $[21]$ (the data are
retrieved on 02.05.2020).

Table \ref{tab:estUSA} represents the estimated model parameters for the
last 5 days of the available data set. Every row in the table represents the
Harris estimate $\tilde{m}_{n}$, as well as it's 95 \% confidence interval ( 
$CI_{l}$ - $CI_{r}$), the proportion of the registered infected individuals $%
\alpha (n)$ and the expected values of the non-confirmed cases $M_{1}$ ( or $%
A_{1}$ for the process with immigration) , based on $n-k$ observations, i.e. 
$Z_{2}(1),...,Z_{2}(n-k);k=0,...,4$.

\begin{table}[tbp]
{\small 
\begin{tabular}{rrcrrr}
$k$ & $\tilde m_{(n-k)}$ & Conf. interval & $\alpha$ & $M_1(n-k)$ & $%
A_1(n-k) $ \\ \hline
4 & 1.0213 & 0.9828 - 1.0598 & 0.38596 & 44651 & 53194 \\ 
3 & 1.0315 & 0.9918 - 1.0712 & 0.38562 & 59410 & 70771 \\ 
2 & 1.0246 & 0.9771 - 1.072 & 0.35219 & 32352 & 38548 \\ 
1 & 1.0545 & 1.0048 - 1.1041 & 0.38064 & 43190 & 51455 \\ 
0 & 1.0286 & 0.9794 - 1.0777 & 0.36665 & 36883 & 43945 \\ 
\end{tabular}
}
\caption{Estimation of the model parameters}
\label{tab:estUSA}
\end{table}

The following figures Figure \ref{fig:Y2} - \ref{fig:A1M1} represent the
parameters of the process for the USA data.

\begin{figure}[H]
\includegraphics[scale=0.3]{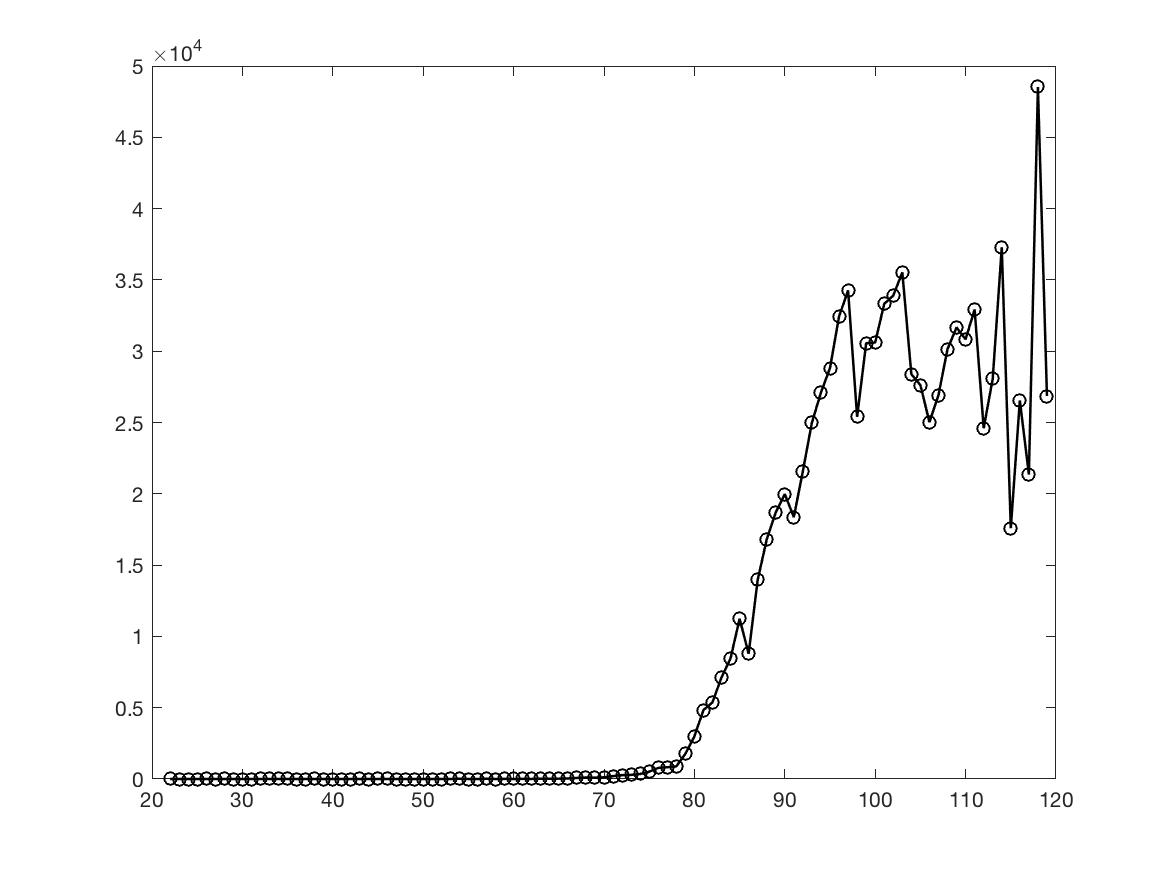}
\caption{Number of the daily reported laboratory-confirmed cases}
\label{fig:Y2}
\end{figure}

Figure \ref{fig:Y2} shows the increments $Y_{2}(n)$ - the number of the
daily reported laboratory-confirmed COVID-19 cases. The related cumulative
values (the number of the total registered $\sum_{i=1}^{n}Y_{2}(i)$) are
presented on Figure \ref{fig:Y2c} exhibiting a strong exponential growth.

\begin{figure}[H]
\includegraphics[scale=0.3]{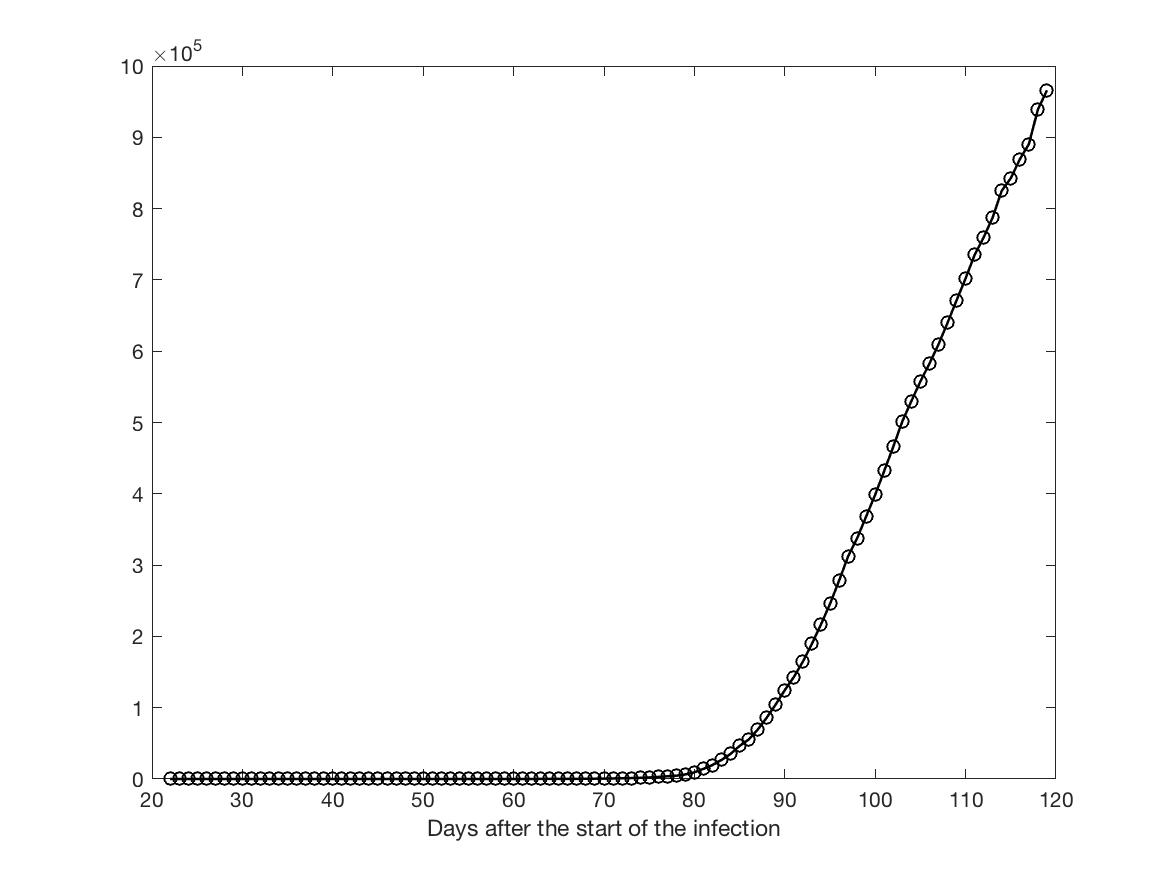}
\caption{Number of the total registered cases}
\label{fig:Y2c}
\end{figure}

A comparison between the Harris type and Lotka-Nagaev type estimators of the
growth rate (the mean value of the newly infected individuals by one
infected individual) can be seen on Figure \ref{fig:H}. After the initially
large estimated values it stabilizes below 1.1, which is determined by the
branching processes theory as a slightly supercritical process. This
corresponds to the exponential growth shown above. The next results shows
that the Harris type estimator has more stable behaviour than the
Lotka-Nagaev type estimator.

\begin{figure}[H]
\includegraphics[scale=0.3]{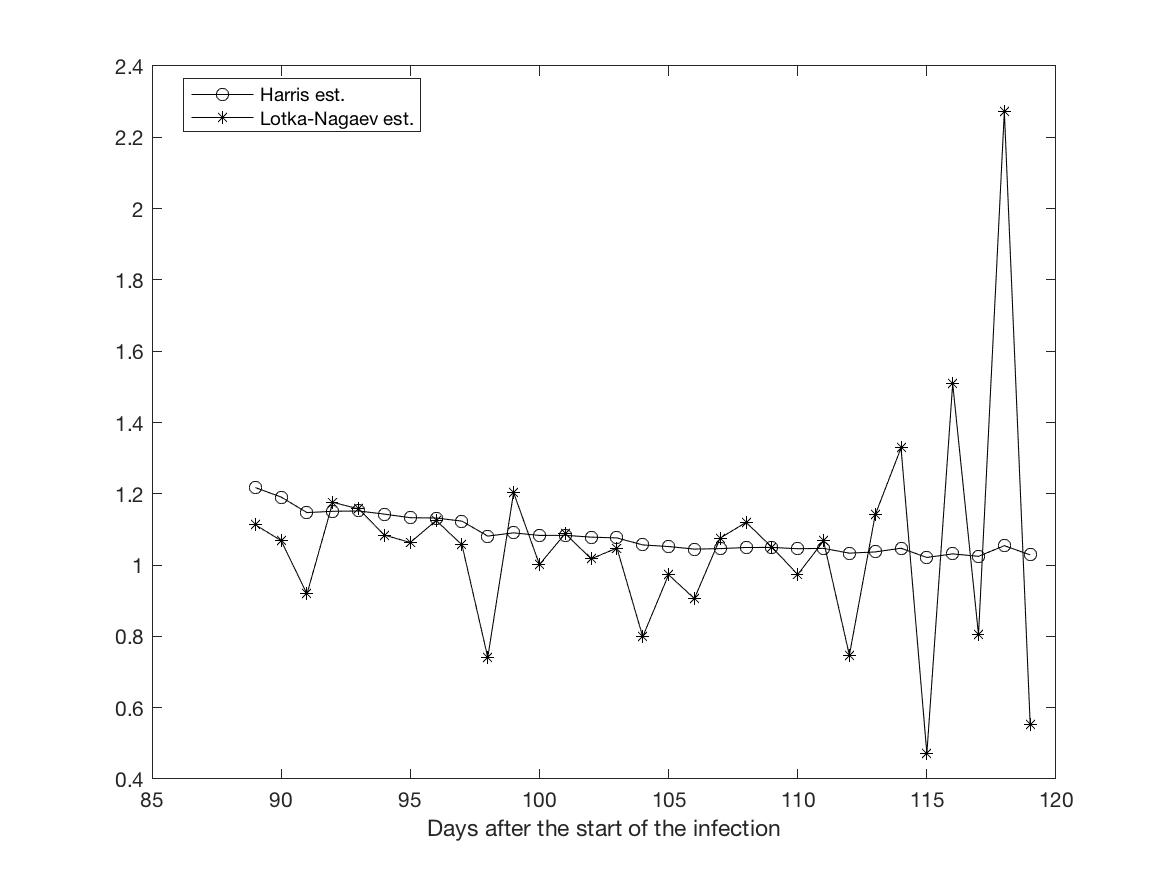}
\caption{The Lotka-Nagaev and the Harris type estimator of the growth rate}
\label{fig:H}
\end{figure}

The estimates of the proportion of the officially registered lab-confirmed
cases among all infected in the population can be seen on Figure \ref{fig:a}%
. During the most recent days their values are approximately 0.8. This means
that nearly 80\% of the infected individuals have been tested, confirmed and
registered.

\begin{figure}[H]
\includegraphics[scale=0.3]{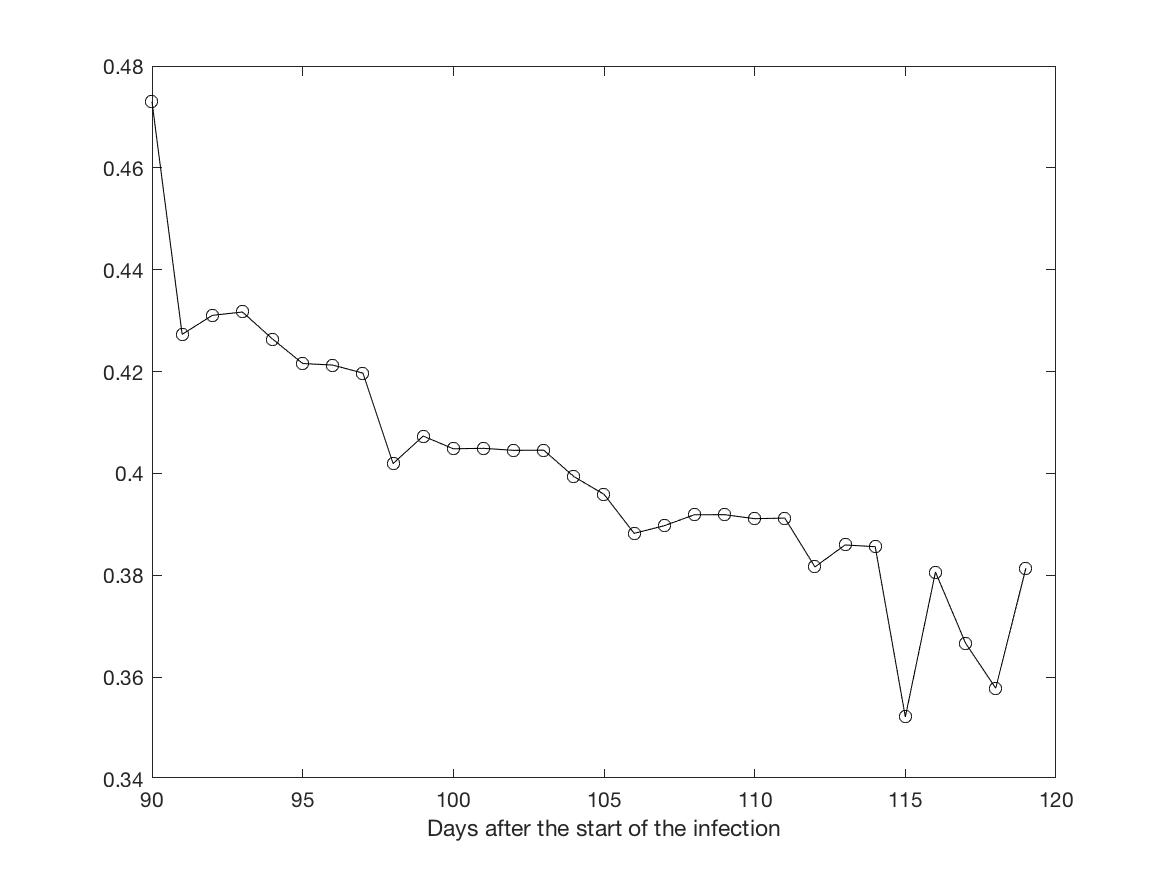}
\caption{Proportion of the officially registered lab-confirmed cases}
\label{fig:a}
\end{figure}

On Figure \ref{fig:M1} the expected number of the nonregistered infected
individuals by days can be seen. The last 5 points on the graph represent
the 95\% confidence interval for the forecast.

The expected number of the nonregistered infected individuals in both cases
- with and without immigration, are compared on Figure \ref{fig:A1M1}.

\begin{figure}[H]
\includegraphics[scale=0.3]{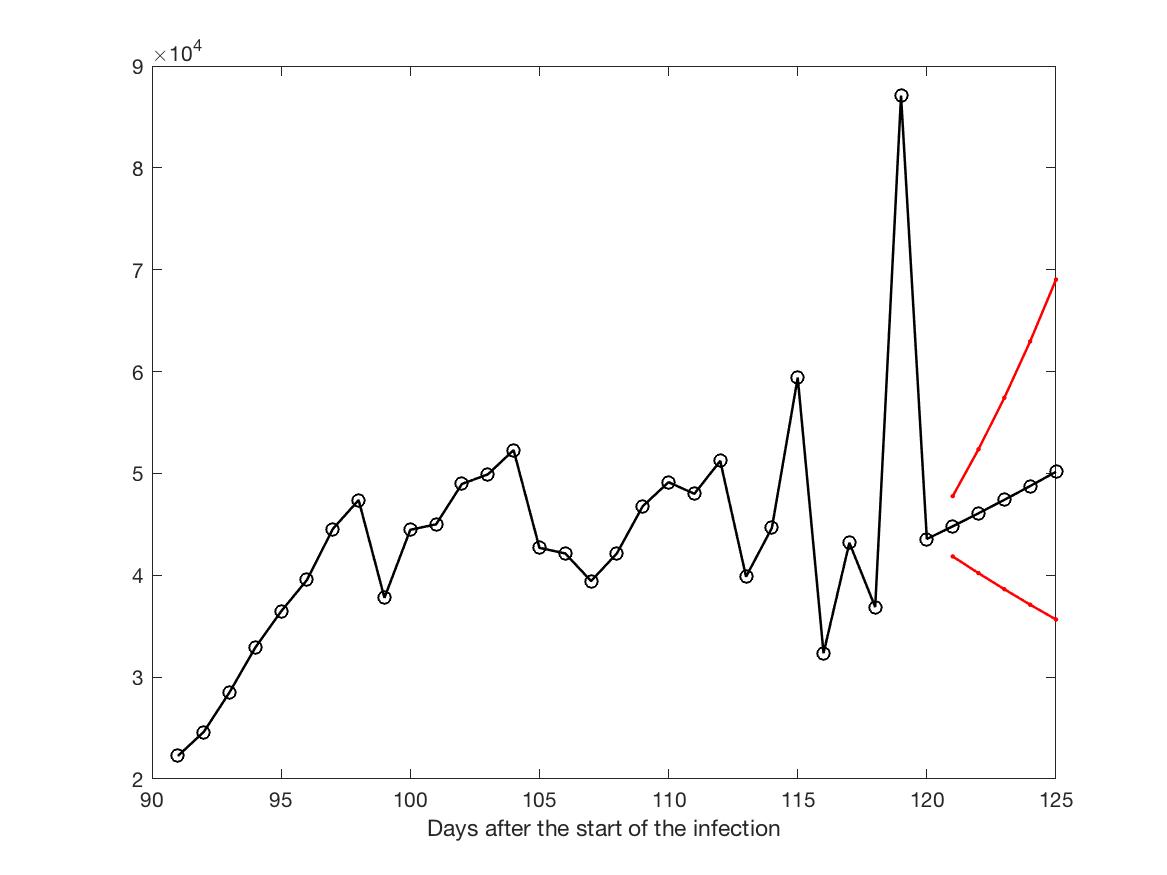}
\caption{Expected number of the nonregistered infected individuals without
immigration}
\label{fig:M1}
\end{figure}

\begin{figure}[H]
\includegraphics[scale=0.3]{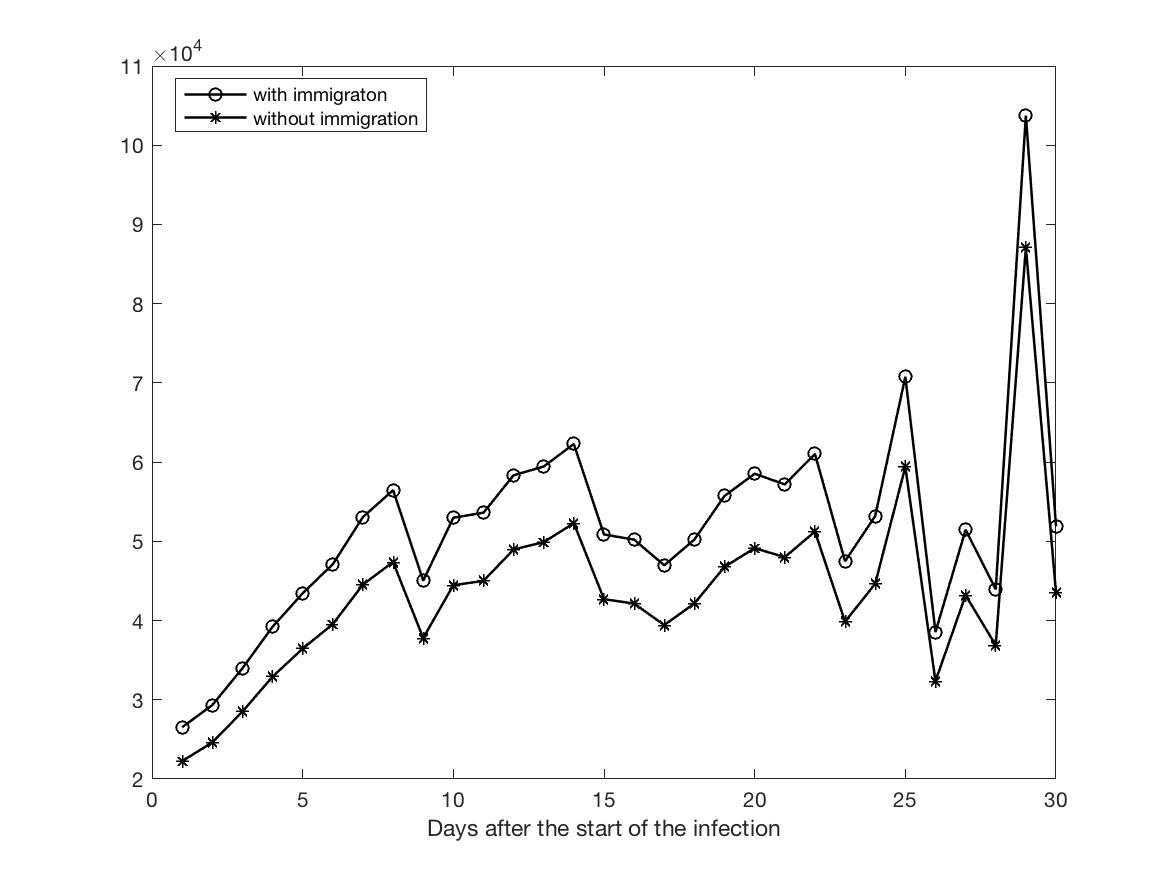}
\caption{Expected number of the nonregistered invected individuals with
immigration}
\label{fig:A1M1}
\end{figure}

Similar results for all countries in the world are available at our
specially constructed site http://ir-statistics.net/covid-19. The data is
provided daily by the European Centre for Disease Prevention and Control $%
[21]$. The results are updated every 24 hours.

The last 5 days results for Italy, Germany, France, Spain and Bulgaria can
be compared on Table \ref{tab:estALL} (the data are retrieved on 02.05.2020).

\begin{table}[tbp]
\begin{center}
{\small 
\begin{tabular}{lrrcrrr}
Country & $k$ & $\tilde m_{(n-k)}$ & Conf. interval & $\alpha$ & $M_1(n-k)$
& $A_1(n-k)$ \\ \hline
Italy & 4 & 1.0183 & 0.9641 - 1.0725 & 0.2189 & 8046 & 8641 \\ 
 & 3 & 1.0141 & 0.9604 - 1.0678 & 0.2397 & 8655 & 9293 \\ 
 & 2 & 1.0159 & 0.9626 - 1.0692 & 0.2608 & 9549 & 10249 \\ 
 & 1 & 1.0122 & 0.9592 - 1.0651 & 0.2332 & 8699 & 9339 \\ 
 & 0 & 1.0119 & 0.9596 - 1.0642 & 0.2458 & 9265 & 9944 \\ 
France & 4 & 1.01550 & 0.8575 - 1.1735 & 0.2589 & 5871 & 6713 \\ 
 & 3 & 1.0138  & 0.8582 - 1.1694 & 0.2828 & 6764 & 7731 \\ 
 & 2 & 1.0147  & 0.8615 - 1.1677 & 0.2430 & 5689 & 6506 \\ 
 & 1 & 1.0125  & 0.8617 - 1.1633 & 0.2303 & 5522 & 6316 \\ 
 & 0 & 1.0037  & 0.8551 - 1.1523 & 0.2365 & 5724 & 6546 \\ 
Spain & 4 & 1.0155 & 0.9340 - 1.0969 & 0.2191 & 11944 & 13471 \\ 
 & 3 & 1.0138 & 0.9334 - 1.0942 & 0.1927 & 11006 & 12413 \\ 
 & 2 & 1.0145 & 0.9352 - 1.0938 & 0.2064 & 11734 & 13234 \\ 
 & 1 & 1.0084 & 0.9302 - 1.0865 & 0.1953 & 11401 & 12858 \\ 
 & 0 & 1.0000 & 0.9229 - 1.0770 & 0.1999 & 11779 & 13284 \\ 
Germany & 4 & 1.0161 & 0.8954 - 1.1368 & 0.1735 & 8499 & 9219 \\ 
& 3 & 1.0158 & 0.8966 - 1.1349 & 0.1944 & 9265 & 10047 \\ 
& 2 & 1.0137 & 0.8963 - 1.1310 & 0.1977 & 9543 & 10348 \\ 
& 1 & 1.0114 & 0.8958 - 1.1270 & 0.1956 & 9605 & 10415 \\ 
& 0 & 1.0066 & 0.8926 - 1.1205 & 0.1825 & 9202 & 9979 \\ 
Bulgaria & 4 & 1.0482 & 0.8672 - 1.2291 & 0.2609 & 40 & 90 \\ 
 & 3 & 1.0693 & 0.8737 - 1.2650 & 0.3328 & 92 & 149 \\ 
 & 2 & 1.0811 & 0.8922 - 1.2701 & 0.3319 & 99 & 156 \\ 
 & 1 & 1.0480 & 0.8637 - 1.2322 & 0.2917 & 177 & 244 \\ 
 & 0 & 1.0409 & 0.8641 - 1.2177 & 0.2610 & 258 & 335 \\ 
\end{tabular}
}
\end{center}
\caption{Estimation of the model parameters}
\label{tab:estALL}
\end{table}

The value of the proportion of the registered infected individuals is
considerably higher in USA and France than in Germany and Italy, while
countries with a longer infection period observe relatively small values of
the Harris estimator of the mean value of the number of the confirmed
infected individuals by one infected individual. Even more, the lower
boundary of the confidence interval for the Harris estimator falls beneath
the value of 1.

Using the same data set one can compare the infection rate for different
countries and regions on the basis of the estimated growth rate. For
example, on Table \ref{tab:estComp}, the 10 countries with lowest and
highest growth rate are shown. Even in the cases where the infection growth
is less than 1, the upper bound of the confidence interval goes above 1,
which states that there is a possibility that the infection growth will
increase in the future.

In the countries, where the infection is growing, the growth rate is
isoclinically above 1, even though the lower boundary of the confidence
interval is less than 1. It is usually due to the small number of observed
infected individuals.

\begin{table}[tbp]
\begin{center}
{\small 
\begin{tabular}{lrc}
Country & $\tilde m_{(n-k)}$ & Conf. interval \\ \hline
Anguilla & 0.3333 & 0.0666 - 0.6000 \\ 
Faroe Islands & 0.6149 & 0.2740 - 0.9559 \\ 
British Virgin Islands & 0.6666 & 0.41718 - 0.9161 \\ 
United States Virgin Islands & 0.6909 & 0.2338 - 1.148 \\ 
Greenland & 0.8181 & 0.2371 - 1.3992 \\ 
Seychelles & 0.8181 & 0.40229 - 1.2341 \\ 
Bhutan & 0.8571 & 0.7513 - 0.9629 \\ 
Mauritania & 0.8571 & 0.43516 - 1.2791 \\ 
Saint Kitts and Nevis & 0.8666 & 0.3245 - 1.4088 \\ 
Saint Lucia & 0.8666 & -0.0322 - 1.7656 \\  \hline
Chad & 1.1250 & 0.5820 - 1.6679 \\ 
Sri Lanka & 1.1320 & 0.8096 - 1.4545 \\ 
Jamaica & 1.1410 & 0.5097 - 1.7722 \\ 
Cape Verde & 1.1556 & 0.3000 - 3.3325 \\ 
Equatorial Guinea & 1.2009 & 0.3000 - 5.1401 \\ 
Ghana & 1.2103 & 0.24158 - 2.1791 \\ 
Palestine & 1.4269 & 0.300 - 3.2266 \\ 
Eswatini & 1.4500 & 0.7094 - 2.1906 \\ 
Maldives & 1.5474 & 0.7273 - 2.3676 \\ 
Ecuador & 2.0316 & 0.4205 - 3.6426 \\ 
\end{tabular}
}
\end{center}
\caption{Comparison of the growth rate}
\label{tab:estComp}
\end{table}

\textbf{5. Concluding remarks.}

First of all the estimation of the mean value of reproduction $m$ allows us
to classify the contamination process as supercritical ($m>1$), critical ($%
m=1$) and subcritical ($m<1$). Recall that both type processes with or
without immigration have exponential growth in the supercritical case $m>1$.
In the critical case $m=1$ and in the subcritical case $m<1$ the asymptotic
behaviour is essentially different. In the critical case the mean value of
the process with immigration grows linearly while for the process without
immigration the mean value is constant. In the subcritical case the mean
value of the immigration process converges to a positive constant but for
the process without immigration the mean value limit is equal to zero.

Finally the estimating of the mean parameter of infection can be considered
as a first stage to construction of a more complicated epidemiological
model. As an example, one can use a branching process with random migration
considered in $[17-19]$ or some other model of controlled branching
processes (see $[8]$). But for a general pandemic model the collaboration
with specialists of epidemiology, mathematics, medicine, microbiology,
molecular biology and informatics is absolutely necessary for the
application of all information about Covid-19 phenomenon.

\textbf{Remark 3. }For more detailed investigation and simulation the
following models can be applied in the considered situation:

$(\mathbf{i)}$\textbf{\ }$h^{\ast
}(s)=q+p_{0}+p_{1}s+p_{2}s^{2}+...+p_{k}s^{k},$ where $q=1-%
\sum_{j=0}^{k}p_{j}$ and $p_{j},j=0,1,...,k,$ can be specially chosen for $%
k=2,3,4,5,6,7,8.$

$(\mathbf{ii)}$ $h^{\ast }(s)=q+p_{0}+\sum_{k=1}^{\infty
}(1-p)p^{k}s^{k}=q+p_{0}+(1-p)ps)/(1-ps),$ where $q+p_{0}=1-p.$ It is
possible to consider also the restricted geometrical distribution up to some 
$k=2,3,4,5,6,7,8.$

$\mathbf{(iii)}$ $h^{\ast }(s)=q+p_{0}+\sum_{k=1}^{\infty }e^{-\lambda }%
\frac{\lambda ^{k}}{k!}s^{k}=q+p_{0}+e^{-\lambda (1-s)}-e^{-\lambda },$
where $q+p_{0}=e^{-\lambda }.$ Similarly it is possible to consider also the
restricted Poisson distribution up to some $k=2,3,4,5,6,7,8.$

Note that the parameters of these distributions can be set in the manner
that $\frac{d}{ds}h^{\ast }(s)|_{s=1}$ is equal to \ $\widehat{m}_{n}$, or $%
\widetilde{m}_{n}$, or $\overline{m}_{n,N}.$ Then with this individual
distributions it is possible to simulate the trajectories of the
non-observed process of infection for further studies.

Additional information, reports and plots, related to this research can be
found on \textbf{http://ir-statistics.net/covid-19}. The presented results
are updated every day following the new data which are provided day by day
from European Centre for Disease Prevention and Control $[21].$

\textbf{Acknowledgements}

The authors would like to express their gratitude to P. Jagers, C. Athreya,
E. Yarovaya, M. Molina, E. Waymire and all the colleagues of the "branching
community" for the useful discussion and suggestions on the first paper $%
[16] $ on Covid-19 topic.

The research was partially supported by the National Scientific Foundation
of Bulgaria at the Ministry of Education and Science, grant No KP-6-H22/3
and by the financial funds allocated to the Sofia University "St. Kliment
Ohridski", grant N: 80-10-116/2020.

\textbf{References}

1. Harris, T.E. The Theory of Branching Processes. Springer, Berlin, 1963.

2. Sevastyanov, B.A. Branching Processes. Nauka, Moscow, 1971. (In Russian).

3. Mode, C.J. Multitype Branching Processes. Elsevier, New York, 1971.

4. Athreya, K.B., P.E. Ney. Branching Processes. Springer, Berlin, 1972.

5. Jagers, P. Branching Processes with Biological Applications. Wiley,
London,1975.

6. Asmussen S., H. Hering. Branching Processes. Birkhauser, Boston,1983.

7. Haccou, P., P. Jagers, V.A. Vatutin. Branching Processes: Variation,
Growth and Extinction of Populations. Cambridge University Press, Cambridge,
2005.

8. Gonzalez, M., I.M. del Puerto, G.P. Yanev. Controlled Branching
Processes. Wiley, London, 2018.

9. Yakovlev, A. Yu., N. M. Yanev. Transient Processes in Cell Proliferation
Kinetics. Lecture Notes in Biomathematics 82, Springer, New York, 1989.

10. Kimmel, M., D.E. Axelrod. Branching Processes in Biology. Springer, New
York, 2002.

11. Guttorp, P. Statistical Inference for Branching Processes. Wiley, New
York, 1991.

12. Yanev, N.M. Statistical inference for branching processes, Ch.7
(143-168) in: Records and Branching processes, Ed. M.Ahsanullah, G.P.Yanev,
Nova Science Publishers, Inc., New York, 2008.

13. Yakovlev, A.Yu., V. K. Stoimenova, N.M. Yanev. Branching processes as
models of progenitor cell populations and estimation of the offspring
distributions. JASA (J.Amer.Stat.Assoc.), 2008, v. 103, no. 484, 1357-1366.

14. Stoimenova, V., D. Atanasov, N. Yanev. Robust estimation and simulation
of branching processes. Proceedings of Bulg. Acad. Sci., T. 57, No. 5, 2004,
19-23.

15. Atanasov D., Stoimenova V., Yanev N. Estimators in Branching Processes
with Immigration. Pliska Stud. Math. Bulgar. \textbf{18}. pp. 19-40. 2007.

16. Yanev, N.M., V. K. Stoimenova, D.V. Atanasov. Stochastic modeling and
estimation of COVID-19 population dynamics. Proceedings of Bulg. Acad. Sci.,
Tom 73, No. 4, 2020. (in press)

17. Yanev, N.M., K.V.Mitov. Critical branching processes with nonhomogeneous
migration. Annals of Probability 13 (1985), 923-933.

18. Yanev,G.P., N.M. Yanev. Critical branching processes with random
migration. In: C.C. Heyde (Editor), Branching Processes (Proceedings of the
First World Congress). Lecture Notes in Statistics, 99, Springer-Verlag, New
York, 1995, 36-46.

19. Yanev, G.P., N.M. Yanev. Branching Processes with two types of
emigration and state-dependent immigration. In: Lecture Notes in Statistics
114, Springer-Verlag, New York, 1996, 216-228.

20. World Health Organization.

https://www.who.int/emergencies/diseases/novel-coronavirus-2019/situation-reports/

21. European Centre for Disease Prevention and Control.

https://opendata.ecdc.europa.eu/covid19/casedistribution/csv/\bigskip

$^{1}$Institute of Mathematics and Informatics, Bulgarian Academy of
Sciences, \newline
yanev@math.bas.bg

$^{2}$Faculty of Mathematics and Informatics, Sofia University,\newline
stoimenova@fmi.uni-sofia.bg

$^{3}$New Bulgarian University,\newline
datanasov@nbu.bg

\end{document}